\documentstyle[preprint,aps,epsf,floats]{revtex}    
\def\bq{\begin{equation}}
\def\eq{\end{equation}}
\def\ba{\begin{eqnarray}}
\def\ea{\end{eqnarray}}
\def\roughly#1{\mathrel{\raise.3ex\hbox{$#1$\kern-.75em\lower1ex\hbox{$\sim$}}}}

\begin{document}
%
%
\preprint{
\font\fortssbx=cmssbx10 scaled \magstep2
\hbox to \hsize{
\hfill$\vtop{   \hbox{\bf MADPH-96-957\hspace{3cm}}
                \hbox{\bf TTP96-38 }
                \hbox{\bf hep-ph/yymmxxx\\}
                \hbox{August 1996}}$ }
}
\title{\vspace*{.5in}
Forward Jet Production at HERA in the Low $x$ Regime\\
       in Next-to-Leading Order}
\author{Erwin Mirkes$^1$ and Dieter Zeppenfeld$^2$\\[3mm]}
\address{$^1$Institut f\"ur Theoretische Teilchenphysik, 
         Universit\"at Karlsruhe,\\ D-76128 Karlsruhe, Germany\\[2mm]}

\address{
$^2$ Department of Physics, University of Wisconsin, Madison, WI 53706, USA}
\maketitle
\begin{abstract}
The production of forward jets of transverse momentum $p_T(j)\approx Q$ and 
large momentum fraction $x_{jet}\gg x$ probes the onset of BFKL dynamics 
at HERA. A full ${\cal O}(\alpha_s^2)$ calculation of the inclusive forward
jet cross section is presented and compared to the expected BFKL
cross section. The kinematical region populated by these events and the scale 
dependence of the fixed order perturbative QCD cross sections are discussed.
\end{abstract}
%
%
\newpage
%
%


Deep-inelastic scattering (DIS) at HERA provides for an ideal place to
probe strong interaction dynamics at small Bjorken $x$. One focus of interest 
has been the rise of the structure function $F_2(x,Q^2)$\cite{lowxF2} for large
$1/x$. One would like to identify the power law growth, $1/x^{\alpha_p-1}$, as 
predicted by the Balitsky-Fadin-Kuraev-Lipatov (BFKL)~\cite{bfkl} evolution 
equation. This evolution equation resums all leading $\alpha_s \ln{1/x}$
terms, as opposed to the more  standard 
Dokshitzer-Gribov-Lipatov-Altarelli-Parisi (DGLAP) equation~\cite{dglap}
which resums all leading $\alpha_s \ln{Q^2}$ terms.
Unfortunately, the  measurement of $F_2$ in the HERA range is probably 
too inclusive to discriminate between BFKL and the conventional DGLAP 
dynamics~\cite{viele}. 

A more sensitive test of BFKL dynamics at small $x$ is expected from deep 
inelastic scattering with a measured forward jet (in the proton direction) 
and $p_T^2(j)\approx Q^2$ \cite{mueller}. The idea is to study DIS events 
which contain an identified jet of longitudinal momentum 
fraction $x_{jet}=p_z(jet)/E_{proton}$ which is large compared to Bjorken $x$. 

\begin{figure}[tbh]
\begin{center}
\hspace*{0in}
\epsffile{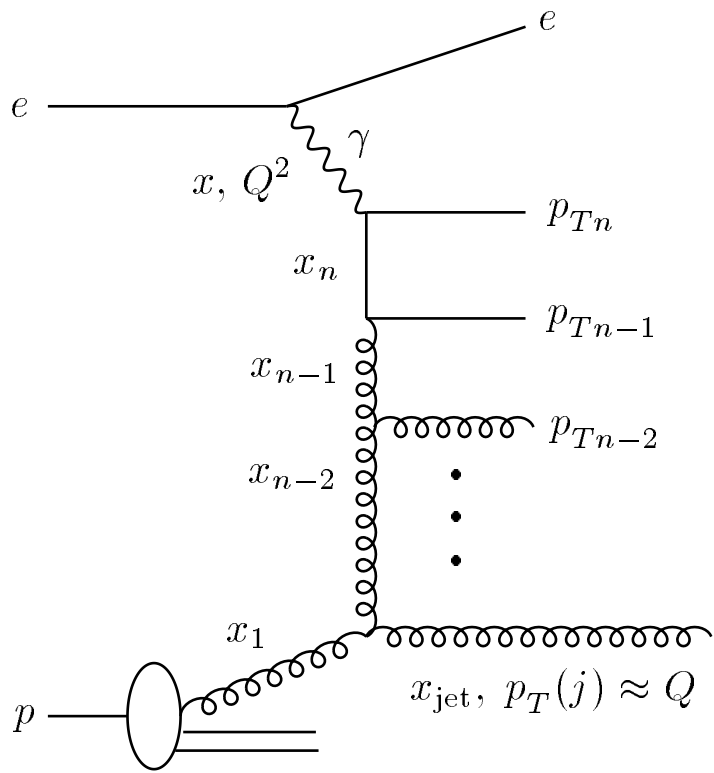}
\vspace{1cm}
\caption{
Gluon ladder diagram contributing to jet production in DIS. The position
and kinematics of the parton which can give rise to the forward jet is 
indicated. 
\label{fig:feyn}
}
\end{center}
\end{figure}

One of the dominant Feynman graphs responsible for the parton evolution
is shown in Fig.~\ref{fig:feyn} for the particular case of a gluon
initiated process. The $x_i$ denote the momentum fractions (relative to the 
incoming proton) of the incident virtual partons and $p_{Ti}$ is the 
transverse momentum of emitted parton $i$. In the axial gauge, such 
``ladder-type'' diagrams with strong ordering in transverse 
momenta, $Q^2\approx p_{Tn}^2 \gg \ldots \gg p_T(j)^2$ but only soft 
ordering for the longitudinal fraction $x_1 > x_2>\ldots> x_n\approx x$ 
are the source of the leading log $Q^2$ contribution. 
In the BFKL approximation transverse momenta are no longer ordered along 
the ladder while there is a strong ordering in the fractional 
momentum $x_n \ll x_{n-1}\ll\ldots \ll x_1\approx x_{jet}$. When
tagging a forward jet with $p_T(j)\approx Q$ this leaves little room for
DGLAP evolution while the condition $x_{jet}\gg x$ leaves BFKL evolution 
active. This leads to an enhancement of the forward jet production cross 
section proportional to $(x_{jet}/x)^{\alpha_P -1}$ over the DGLAP 
expectation.

\setlength{\unitlength}{0.7mm}

Fig.~\ref{fig:feyn} shows that the gluon ladders responsible for 
BFKL evolution first appear at ${\cal O}(\alpha_s^3)$, i.e. for processes with 
at least four scattered partons in the final state. A conventional fixed
order QCD calculation up to ${\cal O}(\alpha_s^2)$ contains no trace of BFKL
dynamics and must be considered a background for its detection; one must search
for an enhancement in the forward jet production cross section above the 
expectation for two- and three-parton final states. 

In this Letter we perform a full next-to-leading order (NLO) analysis
of this ``fixed order'' background. Such an analysis has become possible with
the implementation of QCD radiative corrections to dijet production in DIS in
a fully flexible Monte Carlo program, MEPJET\cite{mepjet}. The program 
integrates the cross sections for two-parton and three-parton final states
numerically and at each phase space point the four-momenta of all final state
particles are available. This allows to implement arbitrary experimental cuts
and arbitrary jet definition schemes. 
Collinear and infrared divergences in the naive three-parton 
cross section are removed by slicing the three parton phase space into a hard
region, where all two-parton invariant masses squared are above a technical
cutoff parameter $s_{min}$, and the remaining soft and/or collinear cross 
section is added to the virtual contributions, yielding a finite two-parton
cross section after splitting off initial state collinear divergences into 
universal crossing functions~\cite{giele}. $s_{min}$-independence of the 
resulting cross section~\cite{mepjet} is a powerful test of the integration 
program. For the forward jet phase space region to be discussed below 
we have verified this independence to better than 1\%, 
which is the statistical accuracy of our Monte Carlo runs.

Numerical results below will be presented both for leading order (LO) and 
NLO simulations. The LO 1-jet and 2-jet results employ the LO parton 
distributions of Gl\"uck, Reya and Vogt~\cite{grv,pdflib} together with the 
one-loop formula for the strong coupling constant. At ${\cal O}(\alpha_s^2)$
all cross sections are determined using the NLO GRV parton distribution 
functions $f(x_1,\mu_F^2)$ and the two loop formula for $\alpha_s(\mu_R^2)$. 
With this procedure the 2-jet inclusive rate at NLO is simply given as the 
sum of the NLO 2-jet and the LO 3-jet exclusive cross sections. 
The value of $\alpha_s$ is matched at the thresholds $\mu_R=m_q$ and the
number of flavors is fixed to $n_f=5$ throughout, {\it i.e.} gluons are 
allowed to split into five flavors of massless quarks.

Unless otherwise stated, both the 
renormalization and the factorization scales are tied to the sum of 
parton $k_T$'s in the Breit frame,
\bq
\mu_R = \mu_F = {1\over 2} \sum_i k_T^B(i) \; ,
\eq
where $(k_T^B(i))^2=2E_i^2(1-\cos\theta_{ip})$. Here $\theta_{ip}$ is the 
angle between the parton and proton directions in the Breit 
frame. $\sum_i k_T^B(i)$ constitutes a natural scale for jet
production in DIS~\cite{rheinsberg} because it interpolates between $Q$,
in the naive parton model limit, and the sum of jet transverse momenta, when
$Q$ becomes negligible. 

We are interested in events with a forward jet with $p_T(j)\approx Q$ 
and $x_{jet}\gg x$ and impose kinematical cuts which closely model the 
H1 selection\cite{deroeck} of such events. Jets are defined in the cone 
scheme (in the laboratory frame) with $\Delta R = 1$ and $|\eta|<3.5$. 
Here $\eta=-\ln\tan(\theta/2)$ denotes the pseudo-rapidity of a jet.
Unless noted otherwise, all jets must have transverse momenta of at least 
4~GeV in both the laboratory and the Breit frames. Events are selected 
which contain a forward jet (denoted ``$j$'') in the pseudo-rapidity 
range $ 1.735< \eta(j)< 2.9$ (corresponding to $6.3^o < \theta(j) < 20^o$) 
and with transverse momentum $p_T^{lab}(j)>5$~GeV. This jet must satisfy
\ba
x_{jet}=p_z(j)/E_p > 0.05\;,  \label{eq:fja} \\
0.5<p_T^2(j)/Q^2<4\; ,\label{eq:fjb}
\ea
in the laboratory frame. The condition $x_{jet}\gg x$ is satisfied by 
requiring 
\bq
x<0.004\; .\label{eq:lowx}
\eq 
Additional selection cuts are $Q^2>8~$GeV$^2$, $0.1 < y < 1$, an energy 
cut of $E(l^\prime)>11$~GeV on the scattered lepton, and a cut on 
its pseudo-rapidity of $ -2.868 < \eta(l^\prime)< -1.735$ 
(corresponding to $160^o < \theta(l^\prime) < 173.5^o$).
The energies of the incoming electron and proton are set to 27.5~GeV
and 820~GeV, respectively.

Numerical results for the multi-jet cross sections are shown in 
Table~\ref{table1}. Without the requirement of a forward jet, the cross 
sections show the typical decrease with increasing jet multiplicity which
is expected in a well-behaved QCD calculation. The 3-jet cross section
in the last column constitutes only about 10\% of the 2-jet cross section
and both rates are sizable. The requirement of a forward 
jet with large longitudinal momentum fraction
($x_{jet}>0.05$) and restricted transverse momentum ($0.5<p_T^2(j)/Q^2<4$)
severely restricts the available phase space. In particular one finds that 
the 1-jet cross section 
vanishes at LO, due to the contradicting $x<0.004$ and $x_{jet}>0.05$ 
requirements: this forward jet kinematics is impossible for one single
massless parton in the final state. 

\begin{table}[thb]
\vspace{3mm}
\caption{Cross sections for $n$-jet events
in DIS at HERA at order $\alpha_s^0$, $\alpha_s$, and $\alpha_s^2$. 
The jet multiplicity includes the forward jet which, when
required, must satisfy $p_T(j)>5$~GeV and the cuts of 
Eq.~(\protect\ref{eq:fja},\protect\ref{eq:fjb}). 
The transverse momenta of additional (non-forward)
jets must only exceed cuts of 4~GeV (first and third column). This 
requirement is replaced by the condition $k_T^B>4$~GeV in the second column.
No $p_T^B$ cut is imposed in the 1-jet case at ${\cal O}(\alpha_s^0)$
and the factorization scale is fixed to $Q$.
See text for further details.
}\label{table1}
\vspace{2mm}
\begin{tabular}{l|cc|c}
        \hspace{0.8cm}
     &  \mbox{with forward jet} &
     &  \mbox{without forward jet } \\
     &  \mbox{$p_T^B,p_T^{lab}>4$~GeV}
     &  \mbox{$k_T^B>4$~GeV}
     &  \mbox{$p_T^B,p_T^{lab}>4$~GeV}\\
\hline\\[-3mm]
\mbox{${\cal O}(\alpha_s^0)$: 1 jet}
                    & 0    pb & 0    pb & 8630 pb    \\
\mbox{\mbox{${\cal O}(\alpha_s)$}: 2 jet }
                    & 18.9 pb & 22.4 pb & 2120 pb    \\
\mbox{${\cal O}(\alpha_s^2)$: 1 jet inclusive} 
                    & 100 pb & 100 pb &           \\
\mbox{\phantom{${\cal O}(\alpha_s^0)$:} 2 jet inclusive} 
                    & 83.8 pb & 98.3 pb & 2400 pb    \\
\mbox{\phantom{${\cal O}(\alpha_s^0)$:} 2 jet exclusive}  
                    & 69.0 pb & 66.8 pb & 2190 pb    \\
\mbox{\phantom{${\cal O}(\alpha_s^0)$:} 3 jet }   
                    & 14.8 pb & 31.5 pb & 210  pb    \\
\end{tabular}
\end{table}

Suppose now that we had performed a full ${\cal O}(\alpha_s^2)$ calculation 
of the DIS cross section, which would contain 3-parton final states at tree 
level, 1-loop corrections to 2-parton final states and 2-loop corrections to
1-parton final states. These 2-loop contributions would vanish identically,
once $x\ll x_{jet}$ is imposed. The remaining 2-parton and 
3-parton differential cross sections, however, and the cancelation of
divergences between them, would be the same as those entering 
a calculation of 2-jet inclusive rates. These elements are already 
implemented in the MEPJET program which, therefore, can be used to determine 
the inclusive forward jet cross section within the 
cuts of Eqs.~(\ref{eq:fja}-\ref{eq:lowx}). At ${\cal O}(\alpha_s^2)$ this 
cross section is obtained from the cross section for 2-jet inclusive events 
by integrating over the full phase space of the additional
jets, without any cuts on their transverse momenta or pseudo-rapidities. 
Numerical results are shown in the third row of Table~\ref{table1}. 

The table exhibits some other remarkable features of forward jet events:
the NLO 2-jet inclusive cross section exceeds the LO 2-jet cross section 
by more than a factor four and the 3-jet rate at ${\cal O}(\alpha_s^2)$ is 
about as large as the 2-jet rate at ${\cal O}(\alpha_s)$. These 
characteristics can be understood in terms of the kinematics of forward 
jet events. Kinematics puts severe constraints on the ``recoil system'', 
the part of the final state in the $\gamma$-parton collision of 
Fig.~\ref{fig:feyn} which is left after taking out the forward jet. 
For $x\ll x_{jet}$, a high invariant mass hadronic system must be 
produced by the photon-parton collision. For small scattering angle but large
energy, $E(j)$, of the forward jet this condition translates into 
\bq\label{eq:kinem}
M^2+2E(j)m_T\;e^{-y} \approx \hat{s}_{\gamma,parton}
                 \roughly> Q^2\left({x_{jet}\over x}-1\right) \gg Q^2\; .
\eq
Here $m_T=\sqrt{M^2+p_T^2}$ and $y$ are the transverse mass and rapidity 
of the partonic recoil system. Eq.~(\ref{eq:kinem}) implies that $m_T$ 
must be large, the more 
so the larger the ratio $x_{jet}/x$. On the other hand, the
transverse momentum, $p_T$, of the recoil system is fixed by momentum 
conservation, $p_T = |{\bf q}_T +{\bf p}_T(j)|$, and the cross section is
largest when the transverse momenta of both the virtual photon, $q_T$, and 
of the forward jet are small. Thus, a large recoil transverse mass is 
most easily achieved by two or more partons which create a subsystem of
large invariant mass, $M$, and some of these partons will manifest 
themselves as fairly hard hadronic jets. 

%
\begin{figure}[hbt]
\vspace*{0.5in}            
\begin{picture}(0,0)(0,0)
\includegraphics{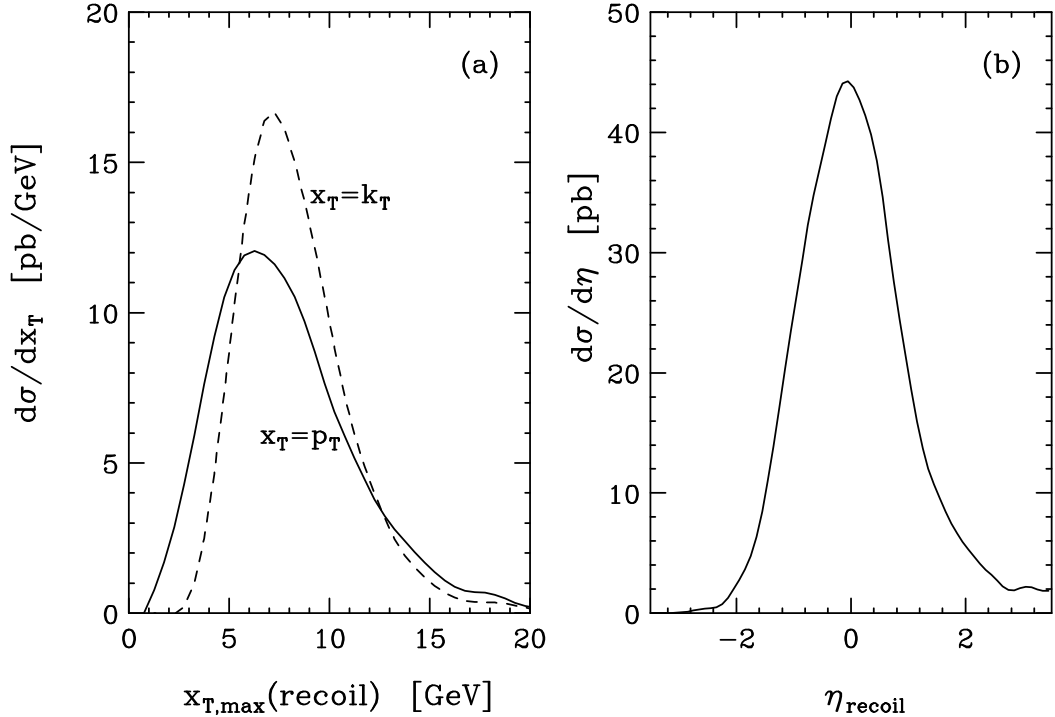}
\end{picture}
\vspace{9.5cm}
\caption{
Characteristics of the highest transverse momentum ``jet'' in the recoil 
system, i.e. excluding the forward jet. Distributions shown are  
(a) $d\sigma/dp_T$ in the lab frame (solid line) and $d\sigma/dk_T$ in the 
Breit frame (dashed line) and (b) the jets pseudo-rapidity distribution in 
the laboratory frame. All distributions are calculated at order $\alpha_s^2$.
Jet transverse momentum cuts have been relaxed to $p_T^{lab},p_T^B>1$~GeV.
\label{fig:recoil}
}
\end{figure}

This fact is demonstrated in 
Fig.~\ref{fig:recoil} where the transverse momentum and the pseudo-rapidity 
distributions of the recoil jet with the 
highest $p_T^{lab}$ are shown, subject only to a nominal requirement 
of $p_T^{lab},p_T^B>1$~GeV. Almost all forward jet events contain at
least one second jet, with $p_T^{lab}\roughly>4$~GeV, which typically 
falls into the central part of the detector. 

In the usual cone scheme final state collinear singularities are 
regulated by the $\Delta R$ separation cut while infrared singularities
and initial state collinear emission are regulated by the $p_T$ cut.
In $\gamma^* p$ collisions the photon virtuality, $Q^2$, eliminates
any collinear singularities for initial state emission in the 
electron direction and therefore a large $k_T$ is as good a criterion 
to define a cluster of hadrons as a jet as its $p_T$.  The dashed line in
Fig.~\ref{fig:recoil}(a) shows the $k_T$ distribution in the Breit frame 
of the recoil jet candidate with the largest $k_T^B$. Basically all 
forward jet events in this NLO analysis possess a recoil ``jet'' 
with $k_T^B>4$~GeV and would thus be classified as 2-jet inclusive events
in a variant of the cone scheme where the $p_T>4$~GeV condition is 
replaced by a $k_T^B>4$~GeV cut. This observation makes intuitively 
clear why we are able to calculate the 1-jet inclusive forward jet cross 
section with a program which calculates the dijet inclusive cross
section at NLO: there exists a jet definition scheme in which all forward
jet events contain at least one additional hard jet. 

\begin{figure}[hbt]            
\begin{picture}(0,0)(0,0)
\includegraphics{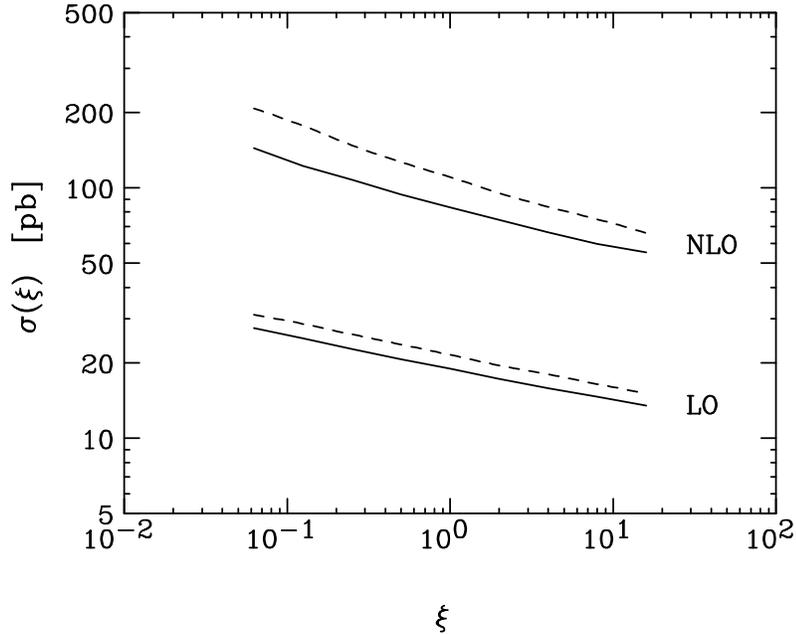}
\end{picture}
\vspace{8cm}
\caption{
Scale dependence of the 2-jet inclusive 
cross section with a forward jet satisfying $x_j>0.05$ 
and $p_{Tj}>5$~GeV in the lab frame (see text for additional cuts). Results 
are shown for $\mu_R^2=\mu_F^2=\xi Q^2$ (dashed lines) 
and $\mu_R^2=\mu_F^2=\xi (0.5\sum k_T)^2$ (solid lines), at LO (lower curves)
and at NLO (upper curves).
\label{fig:scale}
}
\end{figure}

How reliable is the determination of the forward jet cross section at NLO?
The importance of higher order corrections can be estimated by studying 
the dependence of the cross section on the choice of factorization and
renormalization scales, $\mu_F$ and $\mu_R$. Our standard choice is 
$\mu_R^2 = \mu_F^2 = {1\over 4}\left( \sum_i k_T^B(i) \right)^2$. In 
Fig.~\ref{fig:scale} we investigate variations of the 2-jet inclusive 
cross section when changing this scale by a factor,
$\xi$, (solid lines)
\bq
\mu_R^2 = \mu_F^2 = \xi\;{1\over 4} \left( \sum_i k_T^B(i) \right)^2 \; ,
\eq
and we also consider renormalization and factorization scales which are
proportional to the photon virtuality, $Q^2$, (dashed lines)
\bq
\mu_R^2 = \mu_F^2 = \xi\; Q^2  \; .
\eq

%

Two striking features of the forward jet cross section become apparent
in this comparison. The large effective $K$-factor, $K\approx 5$, was
already noted in Table~\ref{table1}. In addition one finds that the scale 
dependence is at least as strong at NLO as at LO. Both features are
closely related. The smallness of the
LO 2-jet compared to the NLO 2-jet inclusive cross section means that
at least three final state partons are required to access the relevant part
of the phase space. This three-parton cross section, however, has only been 
calculated at tree level and is subject to the typical scale uncertainties
of a tree level calculation. Thus, even though we have performed a full 
${\cal O}(\alpha_s^2)$ calculation of the forward jet cross section at HERA,
including all virtual effects, our calculation effectively only gives a LO
estimate of this cross section and large corrections may be expected from
higher order effects, like the gluon ladders in Fig.~\ref{fig:feyn}.

\begin{figure}[t]
\epsfxsize=5.125in
\epsfysize=4.0in
\begin{center}
\hspace*{0in}
\epsffile{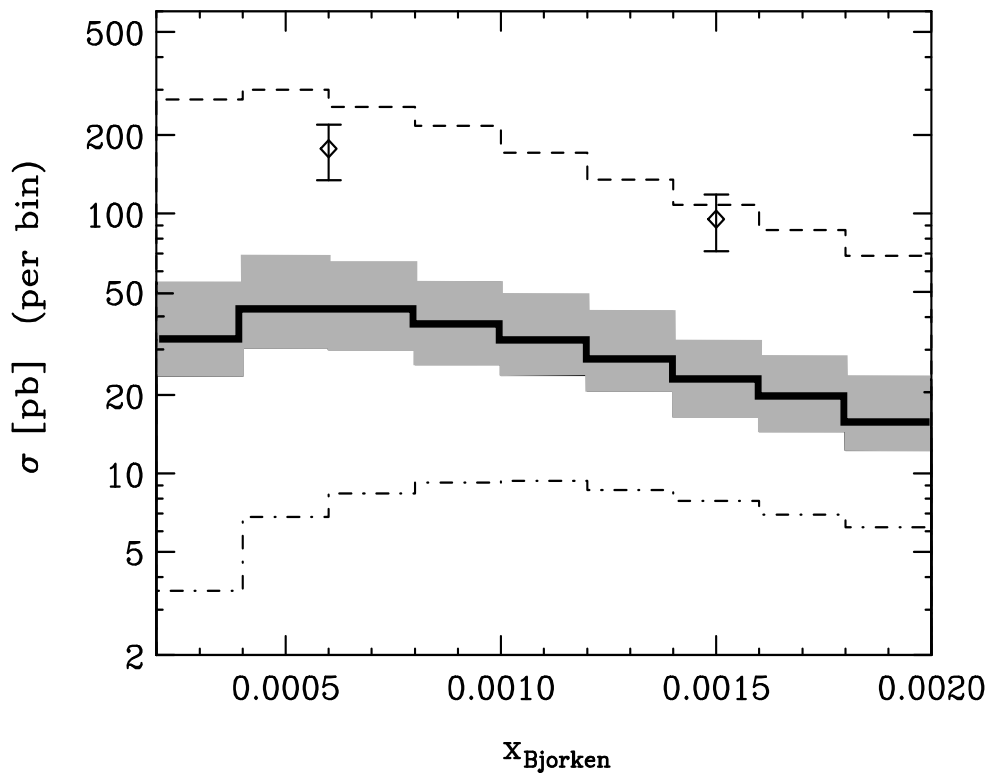}
%
%
\vspace*{0.5cm}
\caption{
Forward jet cross section at HERA as a function of Bjorken $x$ within the H1 
acceptance cuts~\protect\cite{H1result} (see text). 
The solid (dash-dotted) histogram gives the NLO (LO) MEPJET result for  
the scale choice $\mu_R^2=\mu_F^2=\xi(0.5\sum k_T)^2$ with $\xi=1$. 
The shaded area shows the uncertainty of the NLO prediction, corresponding
to a variation of $\xi$ between 0.1 and 10. The BFKL result of 
Bartels et al.~\protect\cite{bartelsH1} is shown as the dashed 
histogram. The two data points with error bars correspond to the H1 
measurement~\protect\cite{H1result}.
\label{fig:h1comp}
}
\vspace*{-0.1in}
\end{center}
\end{figure}

The size of these corrections may be estimated by comparing to BFKL 
calculations or to existing experimental results. The H1 Collaboration has 
published such a measurement which was made during the 1993 HERA run with 
incident electron and proton energies of $E_e=26.7$~GeV 
and $E_p=820$~GeV~\cite{H1result}. The acceptance cuts used for this 
measurement differed somewhat from the ones described before.  Because 
of the lower luminosity in this early HERA run the $x_{jet}$ cut on the 
forward jet was lowered to 0.025 and defined in terms of the jet energy as 
opposed to the longitudinal momentum of the jet in the proton direction,
\bq
x_{jet}=E(j)/E_p > 0.025\;,  \label{eq:H1cuta}
\eq
and the pseudo-rapidity range of the forward jet was chosen slightly 
larger, $ 1.735< \eta(j)< 2.949$ (corresponding to $6^o < \theta(j) < 20^o$).
Scattered electrons were selected with an energy of $E(l^\prime)>12$~GeV and 
in the pseudo-rapidity range $ -2.794 < \eta(l^\prime)< -1.735$ 
(corresponding to $160^o < \theta(l^\prime) < 173^o$). Finally the 
Bjorken-$x$ and $Q^2$ ranges were chosen as $0.0002<x<0.002$ and 
5~GeV$^2<Q^2<100$~GeV$^2$. Within these cuts H1 has measured cross sections
of $709\pm 42\pm 166$~pb for $0.0002<x<0.001$ and $475\pm 39\pm 110$~pb 
for $0.001<x<0.002$. These two data points, normalized to bin sizes of 0.0002,
are shown as diamonds with error bars in Fig.~\ref{fig:h1comp}. Also included
(dashed histogram) is a recent calculation of the BFKL cross 
section~\cite{bartelsH1}.

As shown before, the MEPJET program allows to calculate the full 1-jet 
inclusive forward jet cross section\footnote{We have checked that 
also for the kinematical region considered now almost all forward jet 
events  contain at least one second jet with $p_T^{lab}>4$ GeV 
and $k_T^B>4$ GeV.}
for $x\ll x_{jet}$. The LO result is shown 
as the dash-dotted histogram in Fig.~\ref{fig:h1comp} and the NLO result
is shown as the solid histogram. The shaded area corresponds to a scale 
variation
\bq
\mu_R^2 = \mu_F^2 = \xi\;{1\over 4} \left( \sum_i k_T^B(i) \right)^2 \; ,
\nonumber
\eq
from $\xi=0.1$ to $\xi=10$, and indicates a range of ``reasonable'' 
expectations for the forward jet cross section at ${\cal O}(\alpha_s^2)$. 

While the BFKL results~\cite{bartelsH1} agree well with the H1 data, the 
fixed order perturbative QCD calculations clearly fall well below the 
measured cross section, even when accounting for variations of the 
factorization and renormalization scales. The measured cross section is
a factor 4 above the NLO expectation. The shape of the NLO prediction, 
on the other hand, is perfectly compatible with the H1 results, and not very 
different from the BFKL curve in Fig.~\ref{fig:h1comp}. At LO
a marked shape difference is still observed, which can be traced directly 
to the kinematical arguments given before: according to Eq.~(\ref{eq:kinem})
the transverse mass of the recoil system must increase proportional
to $x_{jet}/x$ and this requires increased transverse momentum of the 
forward jet at LO. Thus, at LO, the expected cross section falls rapidly at 
small $x$, an effect which is avoided when additional partons are available
in the final state to balance the overall transverse momentum.

We conclude that the existing H1 data show evidence for BFKL dynamics in 
forward jet events via an enhancement in the observed forward jet cross 
section above NLO expectations. The variation of the cross section with $x$, 
on the other hand, is perfectly compatible with either BFKL dynamics or 
NLO QCD. Since MEPJET provides a full NLO prediction of the 1-jet inclusive
forward jet cross section for arbitrary cuts and jet definition schemes,
more decisive shape tests may be possible as additional data become available.

\acknowledgements
This research was supported by the University of Wisconsin 
Research Committee with funds granted by the Wisconsin Alumni Research 
Foundation and by the U.~S.~Department of Energy under Grant 
No.~DE-FG02-95ER40896. The work of E.~M. was supported in part  
by DFG Contract Ku 502/5-1.

\newpage

%

\end{document}